\documentclass{epl}

\usepackage{graphicx}% Include figure files
\usepackage{dcolumn}% Align table columns on decimal point
\usepackage{bm}% bold math
\usepackage{epsfig}

\title{Tunable critical current for a vortex pinned by a magnetic dipole }
\author{Gilson Carneiro}
\institute{Instituto de F\'{\i}sica, Universidade Federal do Rio de Janeiro,  
C.P. 68528, 21941-972, Rio de Janeiro-RJ, Brasil}
\pacs{74.25.Sv}{First pacs description} 
\pacs{74.25.Qt}{Second pacs description}

\begin{document}

\maketitle

\begin{abstract}
A simple model for a superconductor with tunable critical current is studied theoretically. The model consists of a thin superconducting film with one vortex interacting with one magnetic dipole, whose magnetic moment is  free to rotate, in the presence of a magnetic field applied parallel to the film surfaces. The pinning potential for the vortex is calculated exactly in the London limit. It is found that, due to the dipole freedom to rotate, the dependence of the pinning potential on the applied field is non-trivial, and allows both the spatial dependence and strength of the pinning potential to be changed by the field. As a consequence, the critical current can be tuned by the applied field. The critical current is obtained numerically as a function of the applied field. Order of magnitude changes in the critical current resulting from changes in the direction and magnitude of the applied field are reported, with discontinuous changes taking place in some cases. Possible application to vortices in low-$T_c$ superconducting films pinned by arrays of magnetic dots are briefly considered.

\end{abstract}

%\section{Section title}
%Paper text.
%Introduction
The pinning of vortices in superconducting films  by arrays of magnetic dipoles placed in the vicinity of the film is a topic that has received a great deal of attention lately. Most of the experimental\cite{rev1} and theoretical\cite{coff,wei,sah,myp,mp,gmc1} work carried out so far deals with arrays of permanent dipoles, that is, dipoles with magnetic moments fixed both in magnitude and direction. A related topic that has received little attention is vortex pinning by arrays of dipoles with magnetic moments free to rotate. The feasibility of fabricating such arrays  has  been demonstrated recently  by Cowburn, et. al.\cite{ckaw}. These authors reported on the magnetic properties of arrays of nanomagnets made of Supermalloy,  each nanomagnet being a thin circular disk of radius $R$. They found that for $R\sim 50-100$nm the magnetic state of each nanomagnet is a single domain one with the magnetization parallel to the disk plane, and that the magnetization can be reoriented by small applied fields. 
One possible source of interest in vortex pinning by freely rotating dipoles is, as demonstrated in this paper, that  the critical current may be tuned by an applied field. This paper studies theoretically in the London limit the interaction between one vortex in a thin superconducting film with one dipole, located outside the film, in the presence of a magnetic field parallel to the film surfaces. The magnetic dipole moment is assumed to be parallel to the film surfaces, to have constant magnitude and  freedom to rotate. Tuning of the critical current is this model results because the interaction between the vortex and the dipole depends on the dipole orientation which, in turn, depends on the applied field. Besides, in a thin film, a magnetic field parallel to the film surfaces has no effect on the vortex in the absence of the dipole. As shown here, this mechanism allows the pinning potential to be changed by the applied field over a wide range.  When a transport current is applied to the film, the magnetic field created by it is  parallel to the film surfaces and  also contributes to the dipole orientation. This makes the pinning potential dependent on the transport current, and has important consequences for the critical current, as shown here. The main new results reported in this paper are:
i) the exact analytic calculation of the pinning potential for one vortex interacting with a freely rotating dipole, and its dependence on the applied field and transport current, ii) the numerical calculation of the critical current for one vortex pinned by the dipole, and its dependence on the magnitude and direction of the applied field. 
%---------------NEW --------------
This paper argues that these results are relevant for vortex pinning by arrays of nanomagnets, similar to those reported in Ref.\cite{ckaw}, placed on top of superconducting films made of homogeneous materials, like most low-$T_c$ ones.  The model is not applicable to layered high-$T_c$ superconducting  films.  
%------------------------------------------------------
 
%Vortex-dipole interaction 

\begin{figure}[h]
\centerline{\includegraphics[scale=0.15]{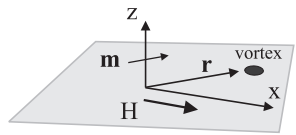}}
\vspace{5mm}
\caption {Superconducting film with one vortex at  ${\bf r}$, a magnetic dipole, ${\bf m}$, at ${\bf r}_{0}=(0,0,z_0)$, and an applied magnetic field, ${\bf H}$, parallel to the film surfaces.}
\label{fig.fig1}
\end{figure}

The calculation of the pinning potential proceeds as follows. The superconductor film is assumed to be planar, with surfaces parallel to each other and to the $x-y$ plane,  isotropic,  characterized by the penetration depth $\lambda$, and of thickness $d\ll \lambda$. A vortex  with vorticity $q$ is  located at position ${\bf r}$, and  the  dipole is  at ${\bf r}_0 =(0,0,z_0>0)$. The dipole moment ${\bf m}$, has constant magnitude, $m$, is oriented parallel to the film surfaces, and is free to rotate in the $x-y$ plane. An uniform magnetic field ${\bf H}$ is applied parallel to the film surfaces. The vortex-dipole system is shown in Fig.\ \ref{fig.fig1}. The total energy in the London limit, neglecting pinning by random material defects,  can be written as \cite{gmc1}  
 \begin{equation}  
E_{T} = - {\bf m}\cdot({\bf b^s_{\perp}} + {\bf H}) +mH
  \label{eq.ett}
  \end{equation} 
where ${\bf b^s_{\perp}}$ is the component parallel to the film surfaces of the field generated by the vortex at the dipole position. The energy $E_{T}$  does not include the vortex self-energy nor the interaction energy of the dipole with the field of the screening current generated by it in the film, because both are independent of the vortex position and dipole orientation.  The constant $mH$ is added for future convenience. 
The parallel component of the vortex field is given in the thin film limit ($d\ll \lambda$) by \cite{bsv} 
 \begin{equation}  
  {\bf b^s_{\perp} }=-q\frac{\phi_0 d}{4\pi \lambda^2}\, \frac{{\bf r}}{r^2}(1-\frac{z_0}{\sqrt{r^2+z^2_0}})
 \, .
  \label{eq.bvt} 
  \end{equation}
This expression is exact for a thin film provided that $r \ll \Lambda=2\lambda^2/d\;$, which is the region of interest here. 
%
%################################################################################# 
%######################################################################
%\end{document}
%
The total energy, $E_{T}$, depends both on the vortex position ${\bf r}$ and on the dipole orientation. The pinning potential for the vortex at zero temperature, denoted by $U_{vm}$, is the total energy for the equilibrium dipole orientation, that is,  for ${\bf m}$ which minimizes $E_{T}$, with the vortex held fixed at ${\bf r}$. Thus, according to Eq.\ (\ref{eq.ett}),  the equilibrium  ${\bf m}$ is  parallel to ${\bf b^s_{\perp}} +{\bf H}$, and the pinning potential is  given by
 \begin{equation}  
U_{vm} = - m\mid {\bf b^s_{\perp}} +{\bf H}\mid + mH \; . 
  \label{eq.ete}
\end{equation}
Note that, by definition, $U_{vm}$ vanishes in the absence of a vortex. According to Eqs.\ (\ref{eq.ete}) and (\ref{eq.bvt}), the spatial dependence of $U_{vm}$  is anisotropic. It depends both on $r$ and on the angle between  ${\bf r}$ and ${\bf H}$. An important consequence of the dipole freedom to rotate is the  non-trivial dependence of $U_{vm}$ on ${\bf H}$ obtained in Eq.\ (\ref{eq.ete}). According to it  ${\bf H}$ plays the role of a handle that controls the  strength and spatial dependence of $U_{vm}$, as will be discussed shortly. The scale for $H$ in Eq.\ (\ref{eq.ete}) is the  vortex field, which  is bound by $b^s_{\perp}\leq b^s_{max}=0.3d/4\pi z_0\times (\phi_0/\lambda^2)$. It is convenient to use  the following natural scales for physical quantities. Energy: $\epsilon_0d$,  where $\epsilon_0=(\phi_0/4\pi\lambda)^2\,$ is the basic scale for energy/length of the superconductor. Magnetic moment:  $\phi_0z_0$. Magnetic field: $\phi_0/\lambda^2$.
%################################################################################# 
\begin{figure}[h]
\centerline{\includegraphics[scale=0.3,clip=]{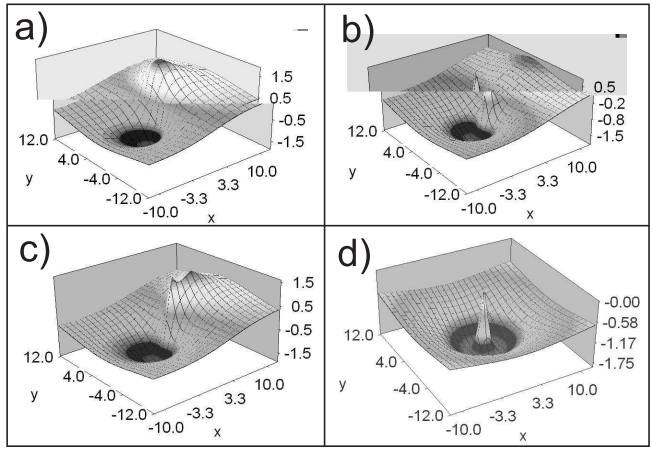}}
\vspace{5mm}
\caption{Spatial dependence of the vortex pinning potential,  $U_{vm}$, (in units of $\epsilon_0 d$) for $q=1$, 
$d=z_0= 2\xi\;, \lambda =10\xi$ ($\xi=$vortex core radius), $m=0.5\phi_0z_0$, and external  field in the $x$-direction.
 a) Permanent dipole. b) $H=0.01\phi_0/\lambda^2$. c) $H=0.02\phi_0/\lambda^2$. d) $H=0$. 
 $x$ and $y$ in units of  $\xi$.}
\label{fig.fig2}
\end{figure}
%######################################################################

For  $H \gg b^s_{max}$ the dipole equilibrium orientation is parallel to ${\bf H}$, and $U_{vm}$ reduces to the pinning potential for a vortex interacting with  a permanent dipole. Assuming that ${\bf H}$ is along the $x$-direction, $U_{vm}=-mb^s_{x}$, which, according to Eq.\ (\ref{eq.bvt}), coincides with the expression obtained in Refs.\cite{wei,sah,mp,gmc1}. In this case $U_{vm}$ is anti-symmetric with respect to both an inversion of the vortex position  (${\bf r} \rightarrow -{\bf r}\; \Longrightarrow \;U_{vm}\rightarrow -U_{vm}$),  and to a change in the sign of the vorticity ( $q\rightarrow -q$). For a vortex ($q>0$),   $U_{vm}$ has a minimum ( maximum ) on the $x$-axis at  $x= - (+)1.3 z_0$, with  minimum (maximum) value  $U_{vm}=-(+)0.3\times 4\pi \epsilon_0 d(m/\phi_0 z_0)$, as shown in Fig.\ \ref{fig.fig2}.a. In general, for   $H \neq 0$   the minimum of $U_{vm}$ occurs when  ${\bf b^s_{\perp}}$ is parallel to ${\bf H}$, that is when the vortex (anti-vortex) is on the negative (positive) $x$-axis. 
In this case, according to  Eq.\ (\ref{eq.ete}), $U_{vm} = - m  b^s_x$. As a consequence, the minimum of $U_{vm}$ for $H\neq 0$ is identical to that for a permanent dipole. However, the spatial dependence of $U_{vm}$ is strongly dependent on H, as shown in Fig.\ \ref{fig.fig2}  for some values of   $H<b^s_{max}$ ($b^s_{max}=0.024\phi_0/\lambda^2$ for the parameters in Fig.\ \ref{fig.fig2}). 
For $H=0$, $U_{vm}$ is given  by $U_{vm} = - m\mid {\bf b^s_{\perp}}\mid $. 
In this case, according to Eq.\ (\ref{eq.bvt}), $U_{vm}$ is the same  for vortices and anti-vortices, has circular 
symmetry, and  is attractive with a repulsive core, as shown in Fig.\ \ref{fig.fig2}.d. The minimum of $U_{vm}$ is  degenerate  on a circle of radius $\,r= 1.3 z_0\,$,  and  has the same minimum value as a permanent dipole ($U_{vm}=-0.3\times4\pi \epsilon_0 d(m/\phi_0 z_0)$).

Now  the critical current, $J_c$, for a single vortex with vorticity $q=1$ is considered. The effect of a  transport
current density, ${\bf J}$,  applied to the film is twofold: it exerts on the vortex a force  
${\bf F}_L=q(\phi_0 d/c){\bf J}\times \hat{{\bf z}}$ and  creates a field at the dipole position 
${\bf H}_J=(2\pi d/c){\bf J}\times \hat{{\bf z}}$, which adds to the external field  and modifies the vortex pinning potential, because $U_{vm}$ is now given by Eq.\ (\ref{eq.ete}) with ${\bf H}$ replaced by the total field ${\bf H}_{T}={\bf H}+{\bf H}_J$. The critical current depends on the relative orientation of  ${\bf J}$ and ${\bf H}$. Here it is assumed that ${\bf J}$ is fixed in the positive $y$-direction, so that both ${\bf F}_L$ and ${\bf H}_J$ are along the positive $x$-direction, and have magnitudes $H_J= 2\pi dJ /c$ and  $F_L=\phi_0dJ/c$, and that ${\bf H}$ points in  a direction that makes an angle $\alpha $  with the positive $x$-axis, that is with ${\bf F}_L$. In this paper $J_c$ is obtained by solving numerically  the equations of motion for the vortex. It is assumed that for ${\bf J}=0$ the vortex is pinned at the absolute minimum of $U_{vm}$, and that $J$ increases very slowly with time. These assumptions ensure that the vortex follows the position of  the minimum of $U_{vm}-  F_L\,x$ as $J$ increase, until $J$ reaches a value for which the  minimum becomes unstable, and the vortex depins. As $J$ increases further, the vortex velocity also increases. The   $J_c$ obtained here corresponds to $J$ for which the vortex velocity reaches a small value chosen for numerical convenience. The obtained $J_c$ is slightly larger than the $J$ for which the minimum becomes unstable. This is analogous to the voltage criterion in  $J_c$ measurements.  The values of $J$ are, of course, limited to $J<J_d$, where $J_d=c\phi_0/(12\sqrt{3}\pi^2\lambda^2\xi)$ is the depairing current, $\xi$ being the vortex core radius. In the results reported next, regions where $J_c>J_d$ are  discussed for the sake of completeness. Now there are two scales for $H$ in $U_{vm}$:  
$b^s_{max}$, as discussed above, and $H_J$. The maximum $H_J$ occurs for $J=J_c$, and can be written as $H_{J_c}=0.031d/\xi(J_c/J_d)(\phi_0/\lambda^2)$. For $d\sim z_0\sim\xi$, these two scales are comparable if $J_c\sim J_d$.  
%################################################################################# 
\begin{figure}[h]
\centerline{\includegraphics[scale=0.30]{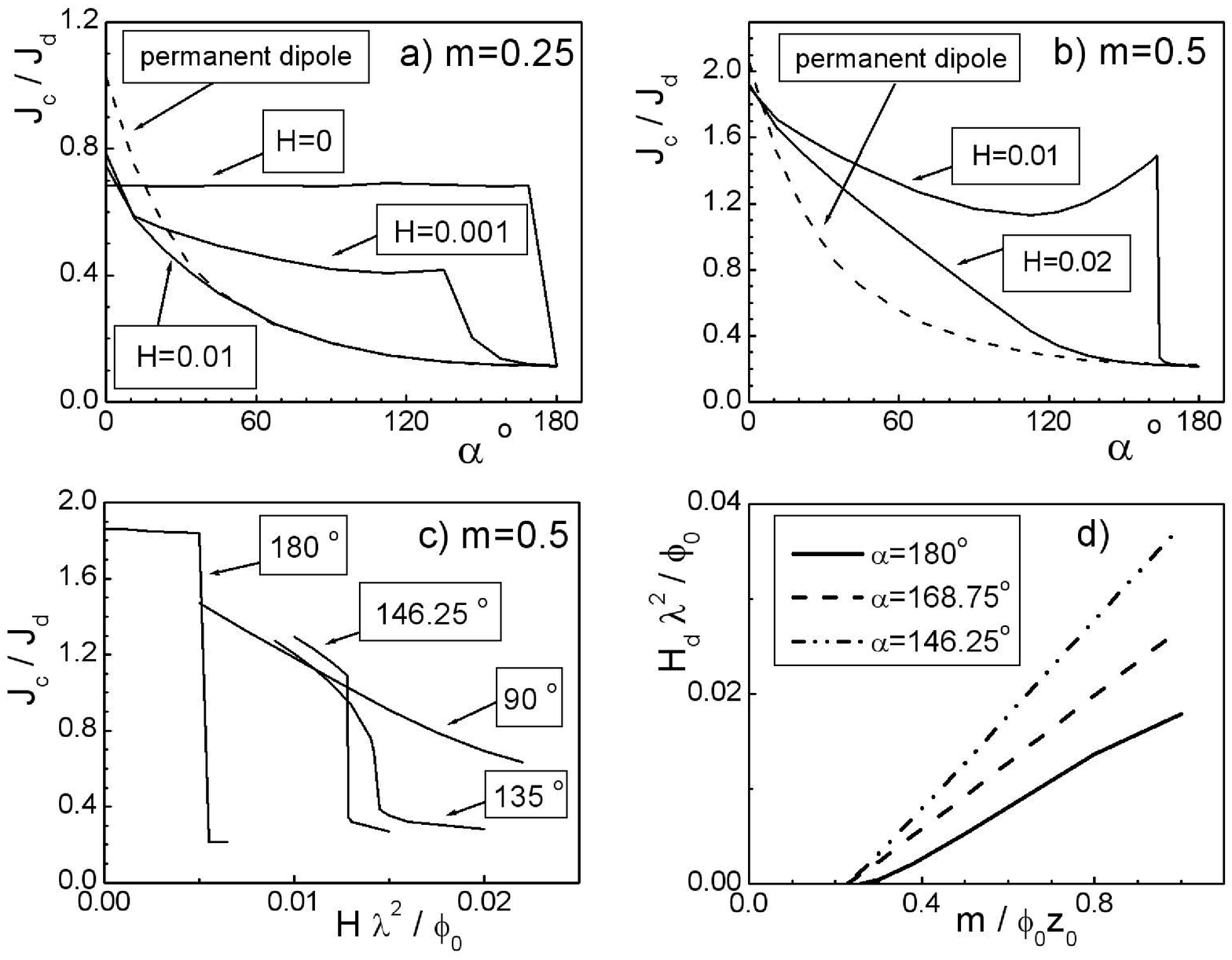}}
\vspace{5mm}
\caption{ Single vortex ($q=1$) critical current for  $\lambda=10\xi, d=z_0= 2\xi$: a) and b) $J_c$ vs. $\alpha$;  c) $J_c$ vs. $H$ for constant $\alpha$, indicated in the boxes; d) discontinuity field $H_d$ vs $m$ . Labels: $m$ in units of $\phi_0z_0$,  $H$ in units of  $\phi_0/\lambda^2$. }
\label{fig.fig3}
\end{figure}
%######################################################################
 
For  $H \gg (b^s_{max},\; H_J)$,  $U_{vm}$  reduces to that for a permanent dipole oriented parallel to ${\bf H}$, that is, with ${\bf m}$ making an angle $\alpha $ with the $x$-axis. In this case,  $U_{vm}$ is independent of $H$ and $J$ and has a spatial dependence like that shown in Fig.\ \ref{fig.fig2}.a rotated by $\alpha$ with respect to the $x$-axis. For $J=0$, the vortex is pinned at the absolute minimum of $U_{vm}$, located at a point in the $x-y$ plane defined in polar coordinates, $(\rho,\theta)$, by $(\rho=1.3z_0,\; \theta=\alpha +\pi)$. The critical current depends on $\alpha$ and $m$, being a linear function of $m$, since $U_{vm}$ is linear in $m$. It is found that $J_c$ depends strongly on $\alpha$, being largest for  
$\alpha=0^o$, and decreasing  smoothly with $\alpha$, as shown in Fig.\ \ref{fig.fig3} a) and b) ( curves labeled {\it permanent dipole}) . This results from the spatial dependence of $U_{vm}$, as can be seen for $\alpha=0^o,\;180^o$, where  the critical current can be estimated analytically,  because the vortex moves only along the $x$ direction as $J$ increases. The result is  $J_c/J_d \simeq 4\, m/\phi_0z_0$   for $\alpha=0^o$, and  $J_c/J_d \simeq 0.4\, m/\phi_0z_0$ for  $\alpha=180^o$.  The origin of this tenfold  difference can be seen in the plot of $U_{vm}$ shown Fig.\ \ref{fig.fig2}.a  The driving force  is parallel  to the  $x$-axis in Fig.\ \ref{fig.fig2}.a for $\alpha=0^o$,  and antiparallel for $\alpha=180^o$. As can be seen in  Fig.\ \ref{fig.fig2}.a, the slope of potential barrier is much steeper in the positive $x$-direction than in the negative one. For other values of $\alpha$ the depinning process is more complicated, because the vortex motion as $J$ increases  is not  confined to the direction of drive. 

For $H$ comparable to or less than $b^s_{max}$ and $H_J$, the equilibrium  orientation  of ${\bf m}$ is no longer fixed, and $J_c$ depends, besides on $\alpha$ and $m$, also on $H$.  Typical results for  $\lambda=10.0\xi$ and $ d=z_0= 2.0\xi$ are shown in Fig.\ \ref{fig.fig3}. The $J_c$ vs. $\alpha$ curves are shown in Fig.\ \ref{fig.fig3}.a  for $m=0.25\phi_0z_0$, and in  Fig.\ \ref{fig.fig3}.b for $m=0.5\phi_0z_0$,  for characteristic  values of $H$. In both cases the $J_c$ vs. $\alpha$ curves differ considerably from those for a permanent  dipole for small 
$H$, being strongly dependent on $H$,  and showing sharp changes in $J_c$ close to $\alpha=180^o$, like those  for  $m=0.25\phi_0z_0$, $H=0.001\phi_0/\lambda^2$ (Fig.\ \ref{fig.fig3}.a) and  $m=0.5\phi_0z_0$, $H=0.01\phi_0/\lambda^2$  (Fig.\ref{fig.fig3}.b) . The curve labeled $H=0$ in Fig.\ \ref{fig.fig3}.a is the limit of the $J_c$ vs. $\alpha$ curve as $H\rightarrow 0$ with $\alpha$ fixed. The strong dependence of $J_c$ on $H$  is even more evident if $J_c$ is plotted as a function of  $H$ for  fixed $\alpha$,  as shown in Fig.\ \ref{fig.fig3}.c for $m=0.5\phi_0z_0$. In this case it is found that for  $\alpha \geq 146.25^o$ the $J_c$ vs. $H$ curves have discontinuities at $H=H_d$, jumping from  $J_c >J_d$ for $H<H_d$ to $J_c \sim 0.2J_d $ for $H>H_d$.  For  $\alpha < 146.25^o$, the dependence of $J_c$ on $H$ is continuous, as illustrated by the curves  for $\alpha=135^o$ and $\alpha=90^o$. For $\alpha=135^o$,  $J_c$ undergoes a rapid change with $H$ around  $H_d=0.014\phi_0/\lambda^2$, whereas for $\alpha=90^o$ the change in $J_c$ with $H$ is much slower.   It is found that the $J_c$ vs. $H$ curves have no discontinuities if $m$ is smaller than  a minimum value  which depends on $\alpha$. As shown in Fig.\ \ref{fig.fig3}.d,  $H_d$ vanishes at the minimum $m$,  and increases above it  essentially linearly with $m$.
%################################################################################# 
\begin{figure}[h]
\centerline{\includegraphics[scale=0.3]{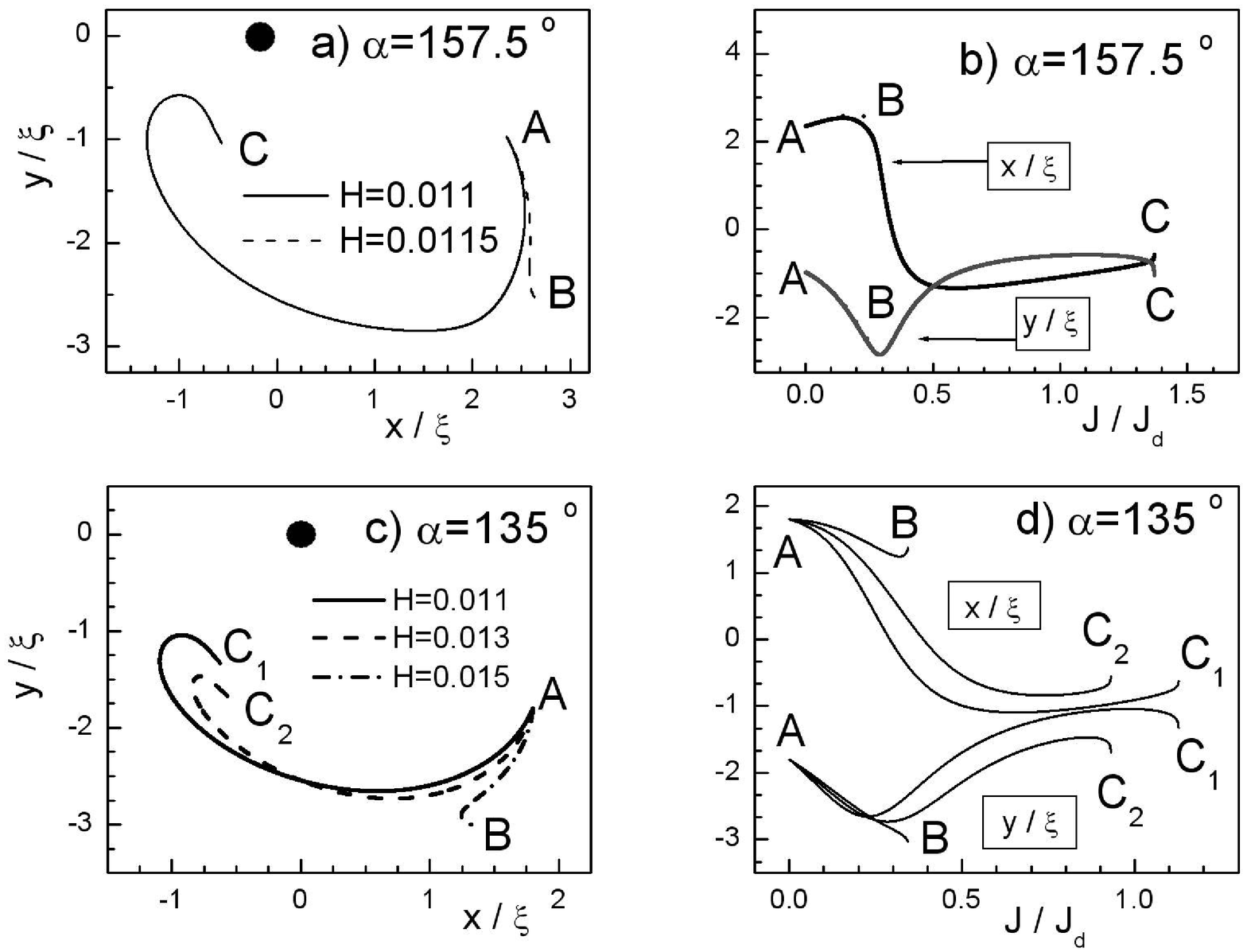}}
%\vspace{5mm}
\caption{Vortex positions with increasing $J$ for $m=0.5\phi_0z_0$, $\lambda=10.0\xi,\, d=z_0= 2.0\xi$. A: Vortex initial position for $J=0$. C: Positions where the vortex depins. a) and c) vortex trajectories. Dot indicates dipole location. b) and d)  $x$ and $y$ coordinates vs. $J$ corresponding to trajectories in a) and c). In d) top curves represent $x/\xi$, bottom curves $y/\xi$. Labels:  $H$ in units of $\phi_0/\lambda^2$. }
\label{fig.fig4}
\end{figure}
%######################################################################

This  complex behavior results from the dependence of $U_{vm}$ on $J$, through ${\bf H}_{T}={\bf H}+{\bf H}_J$, as can be seen by examining how the position the minimum of $U_{vm}- F_L x$, which coincides with the vortex position, changes as 
$J$ increases (Fig.\ \ref{fig.fig4}). For $\alpha=157.5^o;\;H=0.011\phi_0/\lambda^2<H_d$, and  $\alpha=135^o; H=0.011\phi_0/\lambda^2,\, 0.013\phi_0/\lambda^2$, when there are large enhancements  in $J_c$ with respect to the permanent dipole value, the position of the minimum undergoes a large displacement, from the initial one on the right side of the dipole (A in Figs.\ \ref{fig.fig4}.a and \ref{fig.fig4}.c) to the final one, where the minimum becomes unstable, on the left side of the dipole (C in Figs.\ \ref{fig.fig4}.a and \ref{fig.fig4}.c). This is accompanied by a flip in the direction of ${\bf H}_{T}$ from near the negative $x$-axis at $J=0$ to one near the positive $x$-axis when the minimum becomes unstable. The enhancement in $J_c$ results because the vortex is effectively pinned by a permanent dipole oriented at a small angle with the positive $x$-axis.
This  can be seen for  $\alpha=157.5^o;\; H=0.011\phi_0/\lambda^2$ 
( Fig.\ \ref{fig.fig4}.b), which shows that most of the vortex displacement from A to C takes place for $0<J<0.5J_d$. In this interval the direction of ${\bf H}_{T}$ rotates from $157.5^o$ to $12^o$ with the $x$-axis. When the vortex depins, at $J=1.35J_d$, 
${\bf H}_{T}$ points at $3^o$ with the $x$-axis and has magnitude $H_{T}=0.074\phi_0/\lambda^2$.   When there is little or no enhancement in $J_c$ ($\alpha=157.5^o;\;H=0.0115\phi_0/\lambda^2$, and $\alpha=135^o;\; H=0.015\phi_0/\lambda^2$) the position of the minimum undergoes only a small displacement, from A to B in  Figs.\ \ref{fig.fig4}.a and \ref{fig.fig4}.c,  and ${\bf H}_{T}$ points in a direction away from the $x$-axis.

The reason for the discontinuous jumps in $J_c$ is related to way that the stability of the minimum of  $U_{vm}- F_L x$  changes as $J$ increases.  It is found that for $H>H_d$ the minimum becomes unstable twice, whereas for $H<H_d$ it becomes unstable only once.  For $H>H_d$ ( $\alpha=157.5^o;\;H=0.0115\phi_0/\lambda^2$ in Fig.\ \ref{fig.fig4}.a ) the minimum becomes unstable at B, where $J=0.25J_d$. A stable minimum, not shown in Fig.\ \ref{fig.fig4}.a, appears again at a slightly larger value of $J$, and follows a trajectory close to the A-C curve. However, the vortex depins when the minimum becomes unstable for the first time at point B. For $H=0.0115\phi_0/\lambda^2>H_d$ in Fig.\ \ref{fig.fig4}, the minimum only becomes unstable once at point C.

The $J_c$ results described above are believed to be representative of low-$T_c$ superconducting films.  First, the particular set of  parameters   used, $d\sim z_0\sim \xi$, are typical ones for superconducting films with magnetic dots placed on top.  For instance, in the experiments with arrays of magnetic dots with permanent magnetization placed on top of superconducting Nb films, reported in Ref.\cite{pann1},  $d=20nm\sim\xi$. The  magnetic dots  are separated from the film by a thin protective layer of thickness $\sim 20nm$, so that the distance from the magnetic dipole to the film is $z_0\sim \xi$.  Second, since the dependence of $J_c/J_d$ on the model parameters $d$, $z_0$, $m$, $\lambda$, $\xi$, and $H$ is, according to Eqs.\ (\ref{eq.ete}), and  (\ref{eq.bvt}), only through the scaled variables $d/z_0$, $m/\phi_0z_0$, and $H\lambda^2/\phi_0$, many superconducting film-dipole systems are equivalent. 

%-----------------------NEW-----------------------------------------------------------------------------
The London limit is valid for vortices in low-$T_c$ films. However, when a magnetic dipole is placed close to the film, it certainly breaks down if the dipole field  destroys superconductivity locally in the film. Roughly speaking, London theory is valid as long as the maximum dipole field at the film is less than the upper critical field, that is, $m/z^3_0 < \phi_0/(2\pi \xi^2)$, or $m/(\phi_0z_0) < (z_0/\xi)^2/2\pi$. For the values used in the above calculations ($z_0=2 \xi$) this gives $m/(\phi_0z_0) < 0.64$, which is larger than the values  used in this paper. The London limit would be a  better approximation if the present calculations were carried out for larger values of $z_0/\xi$. However, the  results for $J_c/J_d$ would be identical to those described above if $m$ and $d$ were scaled by the same factor as  $z_0/\xi$. For instance, if  $z_0\rightarrow 2z_0$, $J_c/J_d$ would remain the same if  $d\rightarrow 2d$ and $m\rightarrow 2m$, but the upper limit of  $m/\phi_0z_0$ for the  validity of the London approximation would increase by a factor of $4$. The present model also breaks down if $m$ is sufficiently large to create vortices in the film. The threshold value of $m$ for spontaneous vortex creation,  estimated as $m\sim 0.7\phi_0z_0$ using the results of Ref.\cite{gmc1}, is larger tham $m$ used here.  
%----------------------------------------------------------------------------------------

The simple model discussed here is relevant to vortex pinning by arrays of magnetic dots, provided that: i) the dots are sufficiently far apart to neglect dipole-dipole interactions between them, ii) the number of vortices per dot is small enough, so that each dot pins at most one vortex, and  the vortices are far enough apart to neglect vortex-vortex interactions. Unfortunately, there are no experimental results on vortex pinning by magnetic dots with freely rotating magnetic moments to compare the model predictions with. Instead, consider under which conditions  the results described above apply to a system consisting of a typical array of nanomagnets reported in Ref.\cite{ckaw} on top of a thin superconducting film. Assuming that 
$\xi=20nm$, it follows that for  $d=z_0=2\xi$, $\lambda=10\xi$ (as above),  $d=z_0=40nm$,   $\lambda=200nm$, and $\phi_0/\lambda^2=500G$. The value $m=0.5\phi_0z_0$ follows if the disk  radius and thickness are chosen respectively as $R\sim 50nm$ and $t\sim 10nm$, and the disk  magnetization is taken as $M\sim 10^2 \mu_B/(nm)^3$. If the  distance between disks in the array is $a\sim 1\mu m$, the dipole-dipole interaction energy , $E_{dd}\sim m^2/a^3$, is small compared with the vortex pinning potential, $U_{vm}\sim -mb^s_{max}$, since $E_{dd}/U_{vm}\sim 10^{-2}$. The values chosen  for the disk radius and thickness, for the magnetization, and for the distance between disks are  typical of those Ref.\cite{ckaw}. The  results reported above (Fig.\ \ref{fig.fig3}) predict that for $H<b^s_{max}=12G\,$, $J_c$ depends strongly on $H$,  like in Fig.\ \ref{fig.fig3}.c, whereas  for $H>b^s_{max}=12G$, $J_c$ is that for a permanent dipole, and depends only on $\alpha$.  

In conclusion then, this paper demonstrates that the critical current for a vortex in a thin superconducting film pinned by a freely rotating dipole can be tuned by a magnetic field applied parallel to the film surfaces. It is found that tuning takes place for a wide range of fields.  For large fields, when  the dipole moment is stuck in the field direction, the critical current  changes continuously by one order of magnitude when the field is rotated $180^o$, from the direction parallel to the driving force to the direction opposite to it. For fields comparable to the vortex field the critical current is very sensitive to field variations, showing very rapid and even discontinuous changes  by as much as one order of magnitude. It is suggested that the results apply to  experiments  on magnetic dot arrays on top of clean superconducting films. 
\acknowledgments
Research supported in part by the Brazilian agencies CNPq, CAPES, FAPERJ,  and FUJB.

\end{document}